# Activation of telecom emitters in silicon upon ion implantation and ns pulsed laser annealing

G. Andrini[1,2], G. Zanelli[3,4], S. Ditalia Tchernij[3,1,4], E. Corte[3], E. Nieto Hernandez[3], A. Verna[5], M. Cocuzza[5], E. Bernardi[4], S. Virzì[4], P. Traina[4], I.P. Degiovanni[4,1], M. Genovese[4,1], P. Olivero[3,1,4], J. Forneris[3,1,4,*]

[1] *Istituto Nazionale di Fisica Nucleare (INFN), Sezione di Torino, 10125 Italy*
[2] *Dipartimento di Elettronica e Telecomunicazioni, Politecnico di Torino, 10129, Italy*
[3] *Dipartimento di Fisica, Università di Torino, 10125, Italy*
[4] *Istituto Nazionale di Ricerca Metrologica (INRiM), Torino 10135, Italy*
[5] *Dipartimento di Scienza Applicata e Tecnologia, Politecnico di Torino, 10129, Italy*

**Abstract**

The recent demonstration of optically active telecom emitters makes silicon a compelling candidate for solid state quantum photonic platforms. Particularly fabrication of the G center has been demonstrated in carbon-rich silicon upon conventional thermal annealing. However, the high-yield controlled fabrication of these emitters at the wafer-scale still requires the identification of a suitable thermodynamic pathway enabling its activation following ion implantation. Here we demonstrate the activation of G centers in high-purity silicon substrates upon ns pulsed laser annealing. The proposed method enables the non-invasive, localized activation of G centers by the supply of short non-stationary pulses, thus overcoming the limitations of conventional rapid thermal annealing related to the structural metastability of the emitters. A finite-element analysis highlights the strong non-stationarity of the technique, offering radically different defect-engineering capabilities with respect to conventional longer thermal treatments, paving the way to the direct and controlled fabrication of emitters embedded in integrated photonic circuits and waveguides.

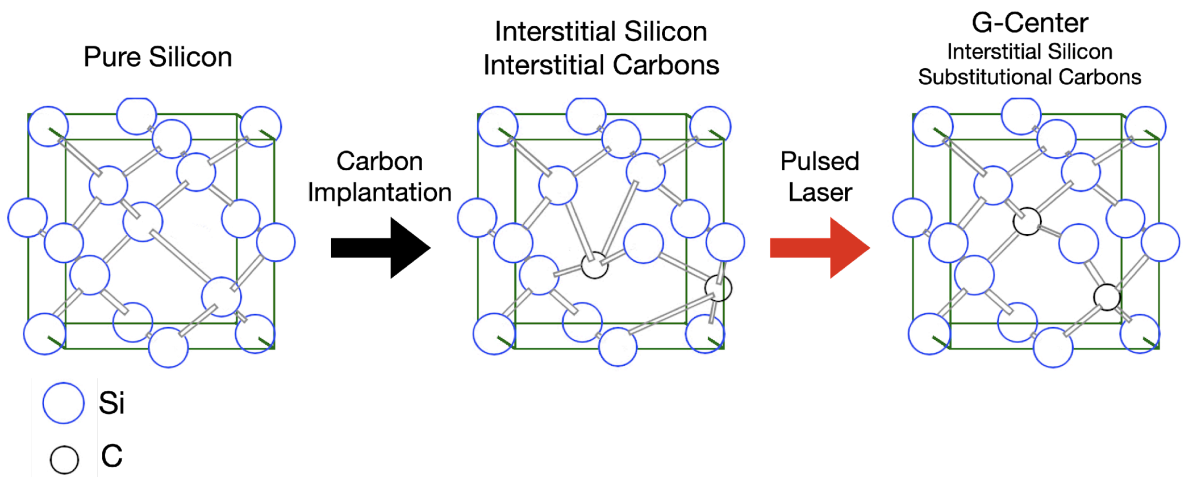

**Introduction**

The G center is a carbon-related point defect in silicon, whose discovery and characterization have surged a strong interest among the scientific community since its recent demonstration as a single-photon emitter.[1,2] In particular, this silicon G center represents an emerging platform in quantum sensing, communication and information processing.[3] [4], due to several promising features, namely: emission in the telecom range (1279 nm);[5,6] availability of a triple-singlet transition enabling optically-detected magnetic resonance protocols;[7] appealing coupling of the defect with nuclear spin and electron spin degrees of freedom.[3,4,8,9]. Furthermore, its availability as a solid-state color center in silicon without the need for articulated homoepitaxial growth processes paves the way towards the development of highly integrable platforms, upon fabrication by means of industry-compatible techniques such as ion implantation. In this context, the technological capability of introducing G centers in high-purity silicon substrates with a high degree of control will be crucial for the fabrication of practical devices. This goal requires mature ion implantation techniques (e.g. deterministic approaches for the delivery of single impurities with high spatial resolution), as well as highly efficient post-implantation processes enabling their optical activation upon conversion into stable lattice defects.[10]

At present, the major obstacle to the systematic implementation of this type of emitter in silicon is represented by its very structural configuration. The point defect has been attributed to a neutrally-charged substitutional dicarbon pair coupled to an interstitial silicon atom.[8,11] This structure, which is rather unusual if compared with that of stable emitters in other classes of materials with diamond-type crystal structure,[12–15] poses questions on the existence of suitable pathways for its consistent and efficient fabrication. A relevant concern, recently highlighted in a theoretical study,[11] is that the configuration of this complex represents a structurally metastable state, in competition with more stable lattice defects that are energetically more accessible during conventional thermal annealing processes. Indeed, the formation of the G center requires the occurrence of two separate processes, namely: the accommodation of a C atoms at a substitutional lattice site and its interaction with a mobile C interstitial [11]. In this respect, it is worth remarking that, while the former process is enhanced by the thermal treatment of the substrate,[9] the latter one is unfavored by annealing treatments and rather promoted by radiation-induced damage within the crystal.[11] This observation is in line with the experimental results reported so far in the literature. The production of the G center has been typically reported in carbon-rich samples, obtained either by Czochralski synthesis[16] or by carbon ion implantation followed by conventional thermal annealing and/or subsequent proton or silicon irradiation.[2,6,17] Notably, the carbon ion implantation step was reported to be unnecessary if the native carbon concentration in the substrate is sufficiently high.[2] [16] These fabrication requirements, accompanied by the necessity of conventional rapid thermal annealing (RTA) treatments (20 s duration)[18] are indicative of a poor efficiency in the formation of G centers under annealing processes characterized by a longer duration.[19] These limitations in the fabrication process lead to two undesirable features. Firstly, the choice of carbon-rich substrates instead of high-purity crystals might result in embedding the fabricated G centers in a defective environment, which might alter their overall emission[20] or increase the background photoluminescence radiation [16]. Secondly, a high-concentration carbon doping of the substrates severely limits to a purely statistical approach the fabrication of the color centers, thus preventing the necessary level of control on their number and position.

In this work we report a new method that allows the spatially resolved and controlled fabrication of G center ensembles, based on ns pulsed focused laser annealing in carbon-implanted silicon substrates. This approach relies on a significantly faster heat transient in the annealing step, with respect to statistical activation approaches based on conventional RTA approaches reported so far.[19]

This work is inspired by the pioneering work carried out by M.S. Skolnick et al.,[21] in which the formation of the G centers was obtained by employing a high power Q-switched laser inducing a melting and subsequent recrystallization of implanted silicon. The dependence of the concentration of the W center in CZ silicon upon pulsed annealing was also further observed by employing a broad infrared source at lasing energy densities insufficient to achieve full recrystallization of the material, thus suggesting a role of thermal heating in the formation of optically active defects [22]. The significant progresses in laser technology have enabled more recently to demonstrate the fabrication of color centers in semiconductors by means of fs laser irradiation[23–25], in which the main role of the optical pulses consists of the introduction of the lattice vacancies necessary for the formation of emitting defects. Similarly to high-power ns lasing, the exploration of fs laser annealing for the fabrication of the G center in silicon was proven to be effective only in case of the crystal melting and subsequent recrystallization. [26]

Conversely, the approach investigated in this work relies on a purely thermal process, replacing conventional annealing treatments while offering the possibility to activate emitters in confined volumes of the target material. By delivering thermal energy to the crystal below the melting threshold, without reaching stationary conditions nor introducing lattice damage, this fast off-equilibrium processing route offers the possibility of strongly enhancing the mobility of the lighter carbon atoms with respect to the silicon interstitials, thus promoting the formation of the G center at the expenses of competing configurations characterized by a higher structural stability. In contrast with other proposed experimental procedures involving surface functionalization through organic molecules,[10] the proposed process offers the activation of G centers with good spatial accuracy, thus enabling in perspective the deterministic activation of individual carbon impurities introduced in the silicon lattice by means of controlled ion implantation protocols. Consequently, this method offers the unique capability of fabricating the G center in higher-purity silicon crystals grown by float-zone (FZ) method, for which RTA is here shown to be ineffective, and in perspective, in ultrapure silicon crystals. In turn, this capability discloses the possibility of fully untapping the potential of the silicon G centers registered to specific nanoscale photonic structures without introducing damage related to local melting of the host material.

**Results**

Our study was performed on a set of samples obtained from a commercial FZ silicon wafer (carbon concentration $<5 \times 10^{14}$ cm$^{-3}$) that was uniformly implanted with 36 keV C$^-$ ions at $2 \times 10^{14}$ cm$^{-2}$ fluence. The PL emission of the implanted sample was investigated with the purpose of comparing the effects of two different post-implantation treatments, namely conventional RTA and ns-pulsed laser annealing.

The photoluminescence (PL) spectra following conventional RTA treatment (20 s duration) are reported in **Fig. 1a** for different temperatures upon normalization to the optical excitation power. Regardless of the processing temperature, all of the RTA treatments resulted in the formation of the sole W center, corresponding to the sharp emission line at 1218 nm and its phonon replicas at higher wavelengths and originating from the extended tri-insterstitial I$_3$-V complex.[27,28] The intensity of the W emission steadily decreases at increasing annealing temperatures, indicating a progressive recovery of the pristine crystal structure. Notably, no spectral features associated with the G center can be identified at any RTA processing stage. Such a result, in apparent contrast with the previous reports on SOI substrates engineering, demonstrates the poor formation of the G center under RTA post-implantation treatment in FZ silicon, and that its formation is arguably limited by the carbon concentration in the silicon substrate. Indeed, the present data acquired from float-zone silicon with low native carbon concentration show that even the introduction of a moderate amount of carbon ions by ion implantation is not sufficient to induce a detectable ensemble of G centers. The interpretation of this experimental evidence is that, despite the fact that most of the carbon impurities evolve into substitutional defects during the annealing, their concentration is not large enough to allow the formation of a small fraction of $(C-Si)_{Si}$ interstitial pairs through the interaction with the silicon interstitials generated during the C implantation. These complexes are necessary for the formation of the G center upon capture at substitutional carbon sites.[11] This interpretation is consistent with the fact that the same implantation and RTA processing parameters adopted in this work have been linked to the formation of ensembles of G centers (although at low densities) in silicon samples characterized by a substantially higher content of substitutional carbon.[19,29] A full understanding of this phenomenon will however require a systematic comparison between different silicon substrates synthesized by different methods in order to enable an optimization of defect engineering procedures.

The laser annealing was performed with a focused ns-pulsed Nd:YAG 532 nm laser (pulse duration: 4 ns; repetition rate: 5 Hz; number of pulses: 5) on 7x7 µm$^2$ square regions of the implanted sample. The sample did not undergo any additional thermal treatment, besides the one under consideration. Confocal PL microscopy was performed using a custom microscope optimized for telecom wavelengths. A typical PL map (488 nm excitation wavelength) acquired from the laser-irradiated sample is shown in **Fig. 1b**. The map exhibits a low luminescence background from the as-implanted sample, within which a series of bright emission squares are clearly distinguishable, corresponding to the laser treated areas. Each square exhibits different emission features depending upon the corresponding laser-processing parameters. A spectral analysis of the PL emission features of the laser-treated regions is summarized in **Figs. 1c** and **1d**.

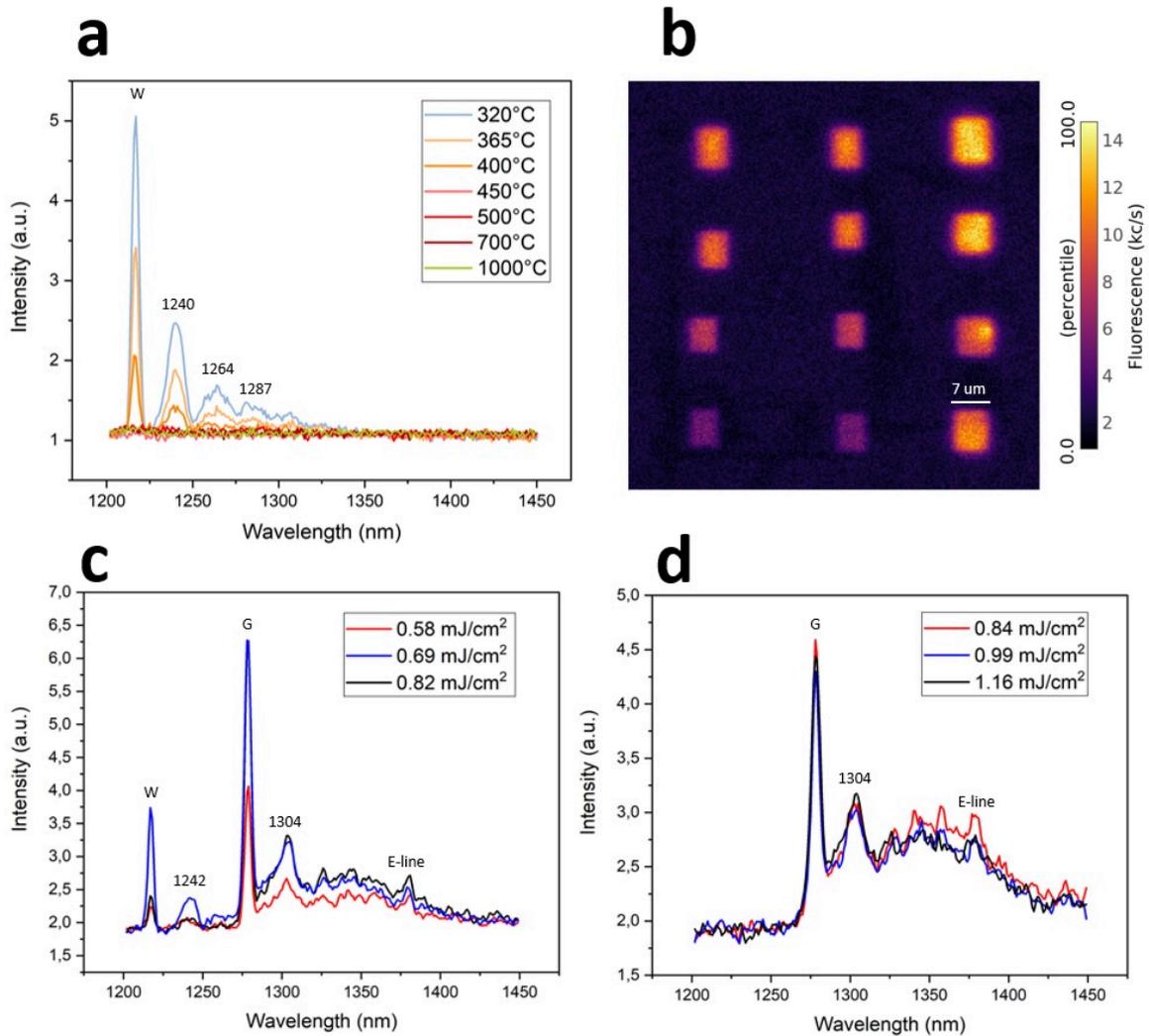

**Figure 1: a)** PL spectra of the sample processed with 20 s rapid thermal annealing process at different temperatures. **b)** typical PL map of the sample processed with localized ns laser annealing. Each bright square corresponds to a spot treated with different lasing parameters. **c)** PL spectra acquired under 488 nm CW excitation from regions lased at 0.58 mJ/cm², 0.69 mJ/cm², 0.82 mJ/cm² **d)** PL spectra acquired in the same conditions from regions lased at 0.84 mJ/cm², 0.99 mJ/cm², and 1.16 mJ/cm². All measurements were performed at T=10 K using a 488 nm excitation and normalized to the optical excitation power.

**Fig. 1c** shows PL spectra from the carbon-ion implanted regions exposed to laser annealing at increasing energy densities (0.58–0.82 mJ/cm²). All of the reported measurements reveal the spectral features of both the W (1218 nm) and G (1279 nm) emitters, whose activation was not achieved by conventional RTA. In the considered spectral range, the intensity of the G center zero-phonon line (ZPL) increases with the lasing energy density; conversely, the W center ZPL shows a clear intensity maximum for the process performed at 0.69 mJ/cm². If the lasing energy density is further increased (**Fig. 1d** shows the PL spectra corresponding to the 0.84-1.16 mJ/cm² range) the W center ZPL disappears, indicating that the defect anneals out under the considered processing conditions. The dependence of the ZPL emission intensities of the G and W centers as a function of laser power is reported in **Figs. 2a** and **2b**, respectively. The peak temperature achieved for each ns pulsed treatment was estimated with numerical simulations (further details in the **Discussion**), and it is reported on the upper horizontal axis, for the sake of comparison with the temperature reached in RTA processing. Both the G and W center ZPL exhibit an initial increase in emission intensity at lower laser power intensities, reaching a maximum value at 0.76 mJ/cm$^2$ and 0.69 mJ/cm$^2$, respectively (these maxima are highlighted by red and green dashed lines in **Fig. 2a-b**). At higher power density values, both the G and W center ZPL emissions exhibit a progressive reduction. It is worth remarking that, in the former case, the emission of the G center does not reduce to a negligible value at the highest (i.e. >1 mJ/cm$^2$) pulse energy densities, but rather reaches a plateau value. An optimal power density range for the formation of the G center with a concurrent attenuation of the PL emission of the W-one is comprised between 0.76 mJ/cm$^2$ and 0.82 mJ/cm$^2$ (corresponding to a local sample temperature of 764-826 °C). In this range, the W center PL is less than half the maximum achieved at 0.69 mJ/cm$^2$, indeed. Finally, **Fig. 2c** shows the "intensity vs temperature" trend for the conventional RTA treatment of the W center. Differently from what is observed for ns laser annealing, the ZPL intensity exhibits a monotonic decrease at increasing annealing temperatures and the center effectively anneals out at temperatures above 450 °C. This process, as already highlighted in **Fig. 1a**, did not result in the formation of optically-active G centers.

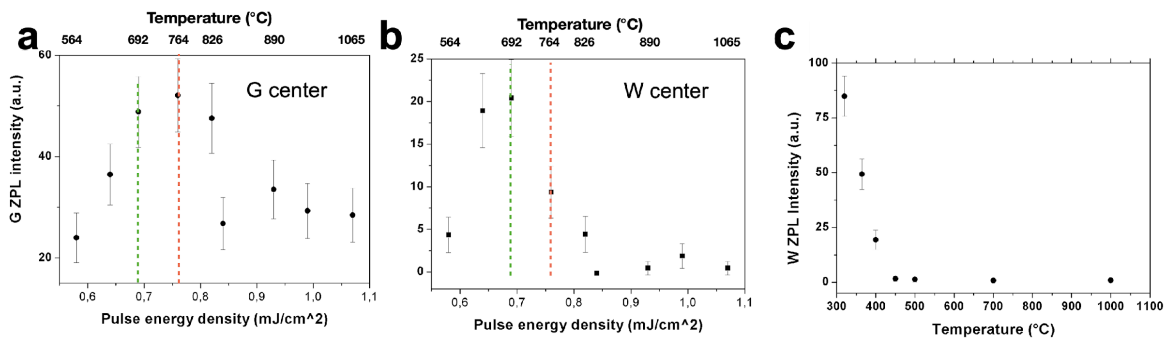

**Figure 2.** ZPL intensity of **a)** G center (1279 nm), and **b)** W-center (1218 nm) as a function of the ns lasing energy density. **c)** ZPL intensity of the W-center as a function of the annealing temperature of the RTA processing (20 s duration). All measurements were performed at T=10 K.

The quality of the emitters was assessed at the ensemble level by investigating the emission linewidth and lifetime, as shown in **Figure 3**. Particularly, Fig. 3a shows the dependence of the ZPL linewidth of the G center on the pulse energy density adopted for the laser annealing process in the 0.5-1.1 mJ/cm$^2$ range. A minimum FWHM value of (0.97 ± 0.05) nm was inferred from a Gaussian fitting at the lowest considered energy density (0.58 mJ/cm$^2$). The FWHM also exhibits an increasing trend with the energy density, reaching a value of (1.10 ± 0.05) nm at 0.76 mJ/cm$^2$ (i.e. the annealing conditions at which the G center ZPL reaches the maximal intensity in the absence of W center emission). These values are in line with what reported for previous measurements in C-implanted silicon both at the ensemble and single-photon level from G centers fabricated by RTA methods[2,6,19,30], thus indicating the suitability of the proposed annealing technique for the fabrication of quantum emitters. Similarly, the lifetime of an ensemble generated by 0.76 mJ/cm$^2$ laser-induced annealing (**Fig. 3b**) was quantified upon 532 nm pulsed laser excitation as (5.7 ± 0.1) ns, in line with the results achieved in SOI silicon undergone C and proton co-implantation and subsequent conventional RTA treatment [6].

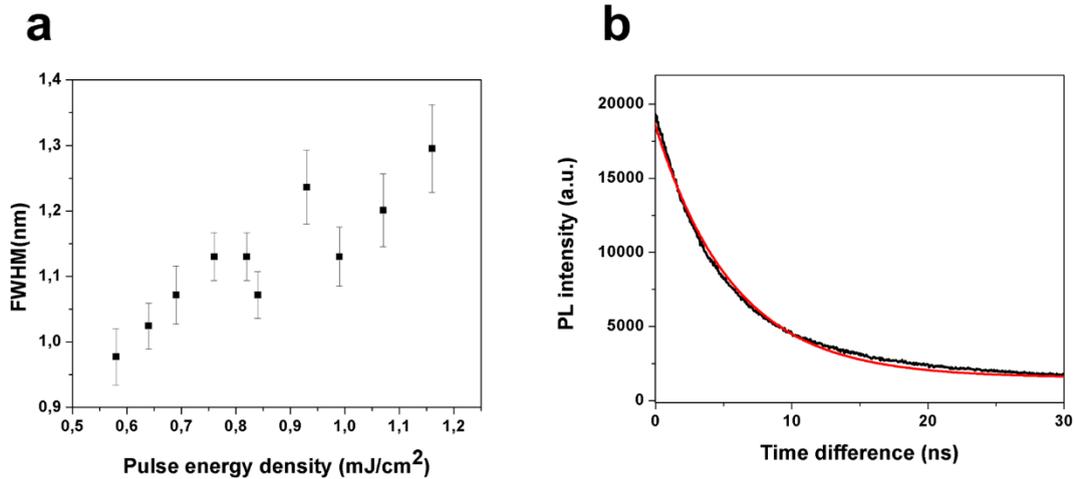

**Figure 3: a)** Linewidth of the G center ZPL as a function of the lasing energy density. **b)** Emission lifetime against a trigger laser pulse acquired from an ensemble of G centers fabricated by 0.76 mJ/cm$^2$ lasing.

The reported results represent the first exploration of laser annealing capabilities in the pulsed ns regime for the activation of optical dopants in solid state. The comparison with the results achieved by conventional RTA (seconds duration), highlights that the light-matter interaction dynamics in the ns regime disclose an effective processing strategy, relying on off-equilibrium temperature transients. Remarkably, while the short duration of the laser pulses allows accessing to meta-stable defective states in a non-equilibrium landscape, the duration of the optical absorption process is at the same time sufficiently long to prevent the introduction of irreversible structural damage to the crystal lattice structure.

Concerning the efficiency achieved in the present experiment, the formation yield, defined as the ratio between the areal density of fabricated optically active emitters centers and the ion fluence[31], was lower bound to 1.4×10$^{-4}$ as estimated by taking into account the detection efficiency of the experimental apparatus (full discussion in the Supplementary Information). For reference, this result is in line with what reported (2×10$^{-4}$) for subsequent implantation steps of C and H ions,[32] while in other works the efficiency is lowered on purpose (e.g. 1.6×10$^{-7}$ for RTA treatments[2]) to decrease the areal density of G centers to the single emitter level.

**Discussion**

A quantitative insight into the peculiar features of the reported laser-based thermal process was made possible by a dedicated finite-element analysis of the heat propagation dynamics in the substrate (**Figure 4**). Firstly, **Figs. 3a-c** show instantaneous temperature maps of the silicon substrate at 4 ns, 40 ns and 400 ns time delays from the delivery of a single 4 ns laser pulse at 1.07 mJ/cm$^2$ energy density. At the end of the lasing pulse (t=4 ns) the maximum temperature (1065 °C, i.e. well below the melting point of silicon) is achieved at the sample surface. The laser-induced heating is highly confined to the irradiated region, with a steep gradient towards environmental conditions occurring over a sub-micron scale. The effect of the annealing is therefore strongly limited to the region that is directly exposed to the laser pulse, as observed experimentally in **Fig. 1b**, and does not involve local recrystallization of silicon.[21,36] Secondly, **Fig. 3d** shows the time evolution of the temperature in correspondence of the center of the laser-irradiated region, at different depths from the sample surface. It is worth remarking that the volume of the region affected by laser-induced heating is also confined in the depth direction. Such localization extends to the first few hundreds nm of the material due to the high (i.e. ~$10^4$ cm$^{-1}$) absorption coefficient of silicon at the 532 nm wavelength under consideration. Temperature does not exceed a 250 °C value at a depth of 3 μm from the surface for the entire duration of the process. Furthermore, the numerical results highlight that the time scale for the whole heat transfer process is shorter than 1 μs. After this transient, the system returns to environmental temperature conditions. The use of a lasing system with 5 Hz repetition rate ensures that any subsequent pulse can be treated as a fully independent annealing process, and thus the process can be reiterated indefinitely at lasing frequencies up to 1 MHz.

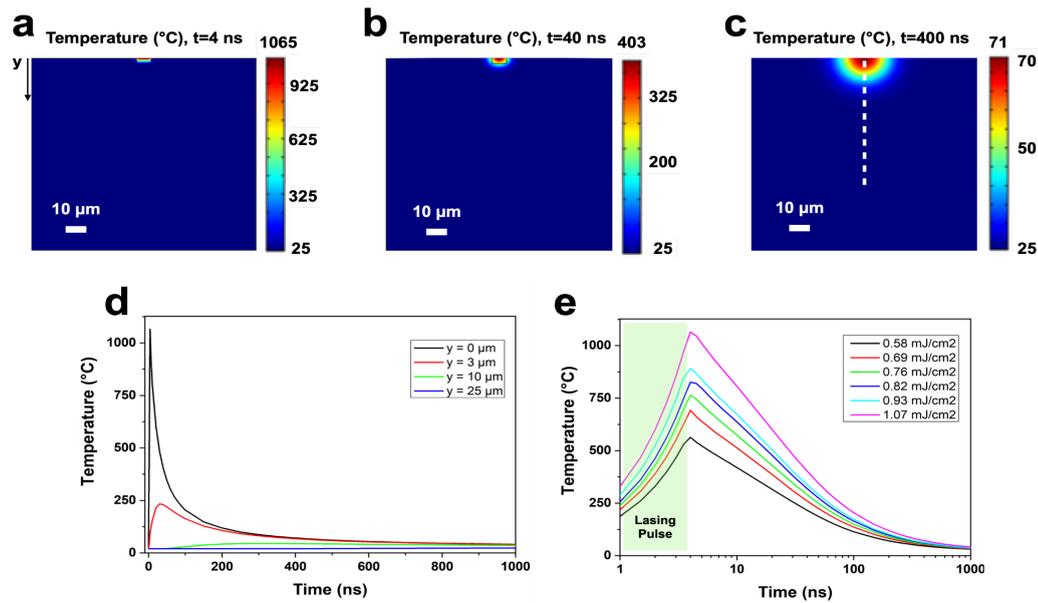

**Figure 4:** Finite element method map of the sample at **a)** t=4 ns, **b)** 40 ns, and **c)** 400 ns after the delivery of a 532 nm laser pulse. The considered energy density is 1.07 mJ/cm$^2$. The 4 ns long laser pulse is switched on at t=0. **d)** Time dependence of the temperature at different depths (z = 0 μm, 3 μm, 10 μm, 25 μm) on the symmetry axis of the system (dashed white line in **Fig. 3c**), evidencing the localized heating of the crystal. **e)** Time evolution of the temperature at the surface (z=0 μm) for different laser energy densities.

Finally, in **Fig. 3e** the temporal evolution of the temperature at the laser-exposed surface is reported for different energy densities. Remarkably, the temperature range (500-1100 °C) spanned in concurrence with the experimentally adopted lasing parameters (0.58–1.07 mJ/cm$^2$) overlaps with the set of temperatures achieved by RTA treatments (**Fig. 1a**). Nevertheless, in the above-mentioned temperature range the ns laser processing is characterized by a lower efficiency at annealing out the W center, thus requiring a simulated temperature of >800 °C (corresponding to an energy density of 0.84 mJ/cm$^2$). This observation further corroborates the interpretation of the ns laser annealing as a strongly non-stationary process resulting in defect-engineering capabilities that are radically different with respect to what can be achieved by conventional RTA treatment.

To conclude, we demonstrated that by means of a pulsed ns laser annealing treatment it is possible to reliably induce the formation of the G center in float-zone silicon. The explored defect-evolution pathway is not accessible under stationary thermal treatments at similar temperatures. This observation is confirmed by the formation (and subsequent deactivation at higher temperatures) of the sole W center by conventional RTA with the process at time scales of few seconds, without the occurrence of any spectral feature associated with the G center. Differently from the previous reports, the defect formation was achieved in a high purity sample in the absence of a high native C carbon concentration, without the need for any post-implantation steps. This result indicates a significantly higher formation of the G center via laser annealing with respect to the currently available methods, and highlights a path towards the conversion into individual emitters of a limited number of C ions achieved by (ideally single-)ion implantation for the controlled fabrication of single-photon sources at telecom wavelengths. Finally, the intrinsically fast (0.2-1.0 µs) nature of the thermal diffusion process enables the indefinite reiteration of the process until the optical activation of the desired impurity is achieved. The localization of the heat absorption to the sole region that is directly exposed to the laser beam, together with the induction of temperatures below the melting point of silicon[21,26] also ensures that the lasing pulses do not alter the optical and structural properties of the surrounding material, where other individual G centers or nanoscale photonic structures might have already been fabricated. We expect that, upon adaptation of the heat diffusion model to material-specific thermal boundaries and different light absorption properties, these considerations can be also applied to SOI devices, with appealing applications in integrated photonics. These features offer an appealing perspective towards the fabrication of large-scale arrays of single emitters, in which high-resolution single-ion delivery is followed by a local (in perspective, down to the optical diffraction limit) annealing process for the conversion of each selected impurity into an individual emitter, provided that single-ion counting at keV energies[37,38] with nanoscale ion beam resolution[16,39] will become consistent techniques routinely available in research labs and industrial manufacturing environments, and upon the assessment of C diffusion length for ns heat transients. Additionally, the fact that the sample temperature is left unchanged by the local annealing originating from individual ns lasing pulses enables the exploitation of in situ feedback PL even at cryogenic temperatures to validate the emitters' formation[40]. The feedback could enable the lasing technique to achieve single defects generation also in substrates with an optimal content of C dopants, without the need for further ion implantation processes. A careful choice of the laser energy density allows to anneal out the W center, thus further confirming that the structural properties of the material are not altered by a disruptive interaction with the laser beam, but that instead the process is also effective at recovering the optical properties of the silicon pristine lattice. All these features are particularly attractive for the non-invasive (i.e. not involving a re-crystallization phase) fabrication of individual emitters registered to specific µm-sized features of a silicon device such as optical waveguides, gratings, or couplers for the development of integrated monolithic devices.[41–44]

**Acknowledgements**
The authors gratefully acknowledge the support of Prof. Ettore Vittone, Physics Department, University of Torino, in the development of the finite-element analysis of the heat propagation dynamics of lased silicon. This work was supported by the following projects: experiment QUANTEP, funded by the 5th National Commission of the Italian National Institute for Nuclear Physics (INFN); Project "Piemonte Quantum Enabling Technologies" (PiQuET), funded by the Piemonte Region within the "Infra-P" scheme (POR-FESR 2014-2020 program of the European Union); 'Training on LASer fabrication and ION implantation of DEFects as quantum emitters' (LasIonDef) project funded by the European Research Council under the 'Marie Skłodowska-Curie Innovative Training Networks' program; "Departments of Excellence" (L. 232/2016), funded by the Italian Ministry of Education, University and Research (MIUR); "Ex post funding of research - 2021" of the University of Torino and QuaFuPhy (bando Trapezio) funded by the "Compagnia di San Paolo". The project 20IND05 (QADeT) and 20FUN02 (PoLight) leading to this publication have received funding from the EMPIR programme co-financed by the Participating States and from the European Union's Horizon 2020 research and innovation programme. 'Intelligent fabrication of QUANTum devices in DIAmond by Laser and Ion Irradiation' (QuantDia) project funded by the Italian Ministry for Instruction, University and Research within the 'FISR 2019' program.

**Methods**

**Sample preparation.** The experiments were conducted on a commercially available float zone n-type silicon wafer (77-95 Ω·cm resistivity) purchased by Vishay Intertechnology. This substrate has been cut up in eight 3x3 mm$^2$ namely identical pieces; each of them was homogeneously implanted with 36 keV $^{12}C^-$ ions at 2·10$^{14}$ cm$^{-2}$ fluence. Two different annealing processes were considered. Conventional rapid thermal annealing was performed using a PID-controlled SSI SOLARIS 150 Rapid Thermal Processing System. Samples #1-7 underwent an annealing for 20 s in N$_2$ atmosphere at 320°C, 365°C, 400°C, 450°C, 500°C, 700°C, 1000°C with a temperature ramp rate of 65°C/s. Pulsed ns flash annealing was performed using a high-power Nd-Yag Q-switched laser source emitting at 532 nm. The source is focused by a primary lens system and collimated by an aperture of 2.75x2.75 mm$^2$. The maximum energy is 0.6 mJ for 4 ns laser pulse duration, thus resulting in a maximum power density of 1.32 MW/cm$^2$. The selected pulse is further focused on the sample by a 20x magnification objective. A repetition rate of 5 Hz was adopted for pulses train delivery.

**IR confocal microscopy**. The PL characterization was performed at cryogenic temperatures by employing a custom fiber-coupled cryogenic confocal microscope. The sample was mounted in a closed-cycle optical cryostation equipped with a vacuum-compatible long-distance 100× air objective (N.A.=0.85). The sample was mounted on a three-axes open-loop nanopositioner. Laser excitation was delivered by a 488 nm CW laser diode, combined with a 700 nm long-pass dichroic mirror and an 800 nm long-pass PL filtering. A multimode optical fiber (core diameter ∅ = 50 μm, N.A.=0.22) was used both as the pinhole of the confocal microscope and as the outcoupling medium for luminescence analysis. The optical fiber fed a InGaAs avalanche detector (MPD PDM-IR / MMF50GI). Spectral analysis was performed by connecting the confocal microscope fiber output to a Horiba iHR320 monochromator, whose output port was in turn fiber-coupled to the PDM-IR detector. The spectral resolution of the system was estimated as ≤4 nm.

**Finite element modeling**. The heat transfer was described by the solution in a two-dimensional model of the time-dependent equation $\rho \cdot c_p \cdot \frac{\partial T}{\partial t} = \nabla \cdot [K \cdot \nabla T] + S$, where $c_p$ is the thermal capacitance and K is the thermal conductivity of isotropic silicon and ρ=2330 kg/m$^3$ represents the mass density. The source term describes the heat absorption from a laser pulse as S=D(x)·P(t)·(1−R)·α·exp(−α·z), where R and α are the reflectivity and absorption coefficient of silicon, assumed to be independent of the temperature and equal to R=0.371 and α=7.69·10$^5$ m$^{-1}$ for 532 nm radiation, respectively[45]. Here, P(t) describes the temporal profile of the laser pulse. Conversely, D(x) describes the profile of the incident laser beam along the x direction parallel to the silicon surface under the assumption of symmetry. The initial temperature was set to 293.15 K.

Two different models were considered to map the substrate temperature. In the first case, a sharp pulse P(t)=$\theta$(t$_0$-t) was considered, where t$_0$=4 ns, and sharp beam D(x)=J/t$_0$·$\theta$(x-x$_0$), J=1.07 mJ/cm$^2$ and x$_0$=7 μm being the pulse energy density and the size of the irradiated square, respectively. K, c$_p$ were assumed to be independent of the temperature.

A second, refined model involved a Gaussian temporal shape of the laser pulse P(t)=(2$\pi\sigma^2$)$^{1/2}$ exp(-(t-t$_0$)$^2$/2$\sigma^2$), where $\sigma_t$= 1.7 ns, and a smoothening of the laser profile at the edges of the irradiated region. In this case, the temperature dependence of the parameters c$_p$, K was set according to tabled data.[46] Boundary conditions were defined to describe thermal convection and irradiation phenomena to occur at the substrate surface and to keep the wafer backside at room temperature. Homogeneous Neumann conditions were applied elsewhere. The simplified model is presented in the main text as it relies on a minimal number of assumptions.

The spatial and temporal distribution of the temperatures did not exhibit significant differences between the two methods, highlighting the small contribution due to the choice of the boundary conditions. The two methods provided a discrepancy in the simulated temperature up to ~90 K at the highest considered pulse energy density, i.e. ~8% of the overall value. This difference is ascribed to the temperature dependence of the thermal capacitance and and conductivity considered in the second model.

## Data availability statement
The data that support the findings of this study are available from the corresponding author J.F., upon reasonable request.